\documentclass{PoS}

\title{Nucleon axial charge in 2+1-flavor dynamical DWF lattice QCD}

\ShortTitle{Nucleon \(g_A\) with 2+1-flavor DWF}

\author{\speaker{Shigemi Ohta}\ 
		for RBC and UKQCD Collaborations\\
        Institute of Particle and Nuclear Studies, KEK, Tsukuba, Ibaraki 3050801 Japan\\
        Department of Particle and Nuclear Studies, SOKENDAI, Hayama, Kanagawa 2400193, Japan\\
        RIKEN-BNL Research Center, BNL, Upton, NY 11973, USA\\
        E-mail: \email{shigemi.ohta@kek.jp}}

\abstract{
The current status of some nucleon isovector observables, the vector charge, \(g_V\), axial charge, \(g_A\), quark momentum fraction, \(\langle x \rangle_{u-d}\), and  quark helicity fraction, \(\langle x \rangle_{\Delta u - \Delta d}\), calculated using recent RBC/UKQCD 2+1-flavor dynamical domain-wall fermions (DWF) lattice QCD ensembles are reported:
with Iwasaki gauge action at inverse lattice spacing, \(a^{-1}\), of about 1.7 GeV, linear lattice extent, \(L\), of about 2.7 fm, pion mass, \(m_\pi\), of about 420 and 330 MeV, and with Iwasaki\(\times\)DSDR gauge action at \(a^{-1}\) of about 1.4 GeV, \(L\) of about 4.6 fm, and \(m_\pi\) of about 250 and 170 MeV.
The calculations have been refined with enhanced statistics, in particular through successful application of the all-mode-averaging (AMA) technique for the 170- and 330-MeV ensembles.
As a result, the precision agreement seen in the charge ratio, \(g_A/g_V\), for 420-MeV and 250-MeV ensembles that share the finite-size scaling parameter \(m_\pi L\) of about 5.8 is more significant with new values of 1.17(2)  and 1.18(4) respectively.
We also studied the dependence on the source-sink separation in the lightest ensemble of 170-MeV, by comparing the cases with the separation of about 1.0 and 1.3 fm and did not see any dependence: contamination from the excited states are well under control in our choice of source and sink smearing.
The axial charge, \(g_A\) and the ratio, \(g_A/g_V\), shows a long-range autocorrelation that extends the entire range of configurations that were so far analyzed, almost 700 hybrid Molecular Dynamics time, in the lightest ensemble of \(m_\pi=170\) MeV.
The other observables do not show any autocorrelation with the interval of 16 trajectories.

\vspace{-197mm}\parbox{\textwidth}{\flushright\large\rm \hfill KEK-TH-xxxx, RBRC-1041}\vspace{194mm}
}

\FullConference{31st International Symposium on Lattice Field Theory - LATTICE 2013\\
		July 29 - August 3, 2013\\
		Mainz, Germany}

\begin{document}

\section{Introduction}

RIKEN-BNL-Columbia (RBC) \cite{Sasaki:2003jh,Orginos:2005uy,Lin:2008uz} and RBC and UKQCD \cite{Yamazaki:2008py,Yamazaki:2009zq,Aoki:2010xg} collaborations have been investigating nucleon structure using domain-wall fermions (DWF).
Through these calculations we have identified persistent deficit in the ratio, \(g_A/g_V\), of isovector axial charge, \(g_A\), to vector charge, \(g_V\), that grew more serious as we moved from quenched \cite{Sasaki:2003jh} to dynamical \cite{Lin:2008uz} ensembles and set pion mass lighter \cite{Yamazaki:2008py}.
The status of our investigations as of Lattice 2012 are summarized in our proceedings \cite{Lin:2012nv,Lin:2012jz}.
We used four recent RBC+UKQCD ensembles \cite{Aoki:2010dy,Arthur:2012opa}:
\begin{enumerate}
\item Iwasaki gauge action, inverse lattice spacing, \(a^{-1}\), of about 1.7 GeV, lattice linear extent, \(L\), of about 2.7 fm, pion mass, \(m_\pi\) of about 420 MeV,
\item same as the above except \(m_\pi\) of about 330 MeV,
\item Iwasaki \(\times\) dislocation suppressing determinant ratio (DSDR) gauge action, \(a^{-1}\) of about 1.4 GeV, \(L\) of about 4.6 fm, \(m_\pi\) of about 250 MeV,
\item same as the above except \(m_\pi\) of about 170 MeV.
\end{enumerate}
We saw the ratio, \(g_A/g_V\), from the ensembles 1 and 3 that share about the same finite-size scaling parameter, \(m_\pi L\), the product of calculated pion mass, \(m_\pi\), and lattice linear extent, \(L\), of about 5.8, agree well, 1.19(4) and 1.15(5), with each other.
We also saw the statistical error grows rapidly as we set the pion mass lighter.
During the past year we enhanced our statistics:
for the ensembles 1 and 3 we almost doubled our statistics by adding more source positions in time, and for the ensembles 2 and 4 we now use the all-mode-average (AMA) technique \cite{Blum:2012my,Blum:2012uh}.

\section{Pion-mass dependence}

\begin{figure}
\includegraphics[width=0.48\textwidth]{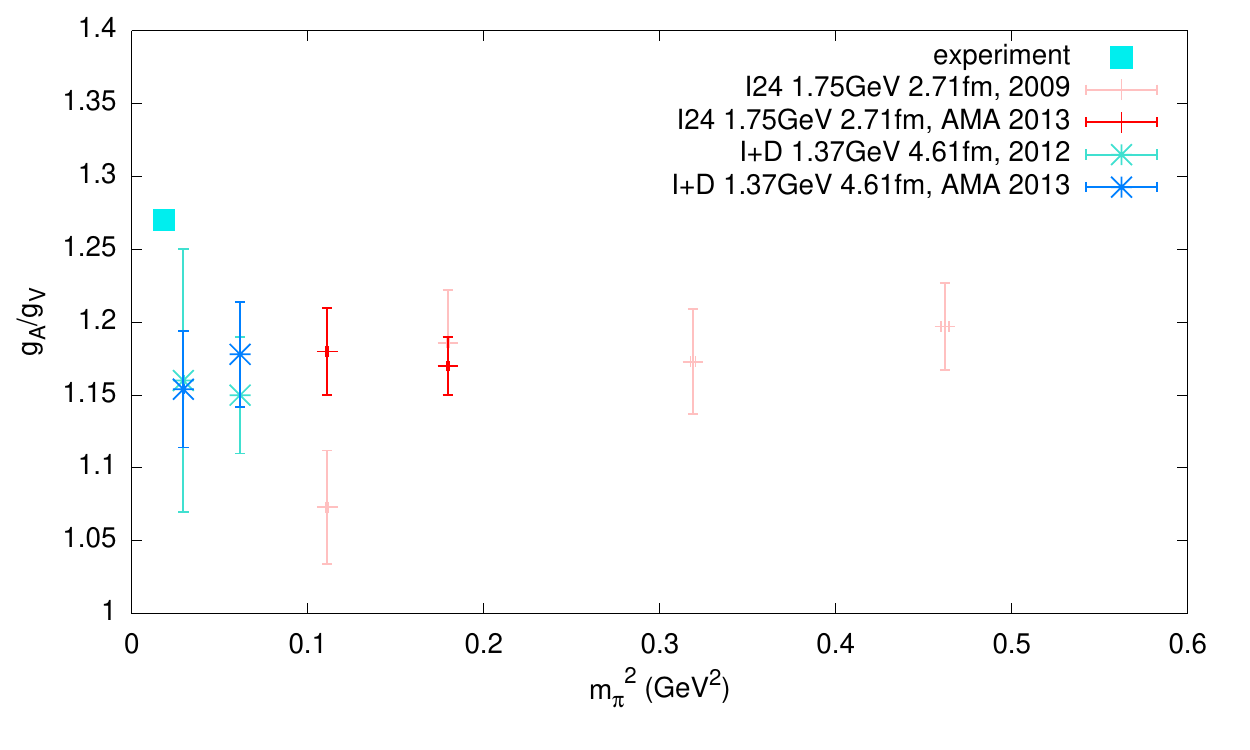}
\includegraphics[width=0.48\textwidth]{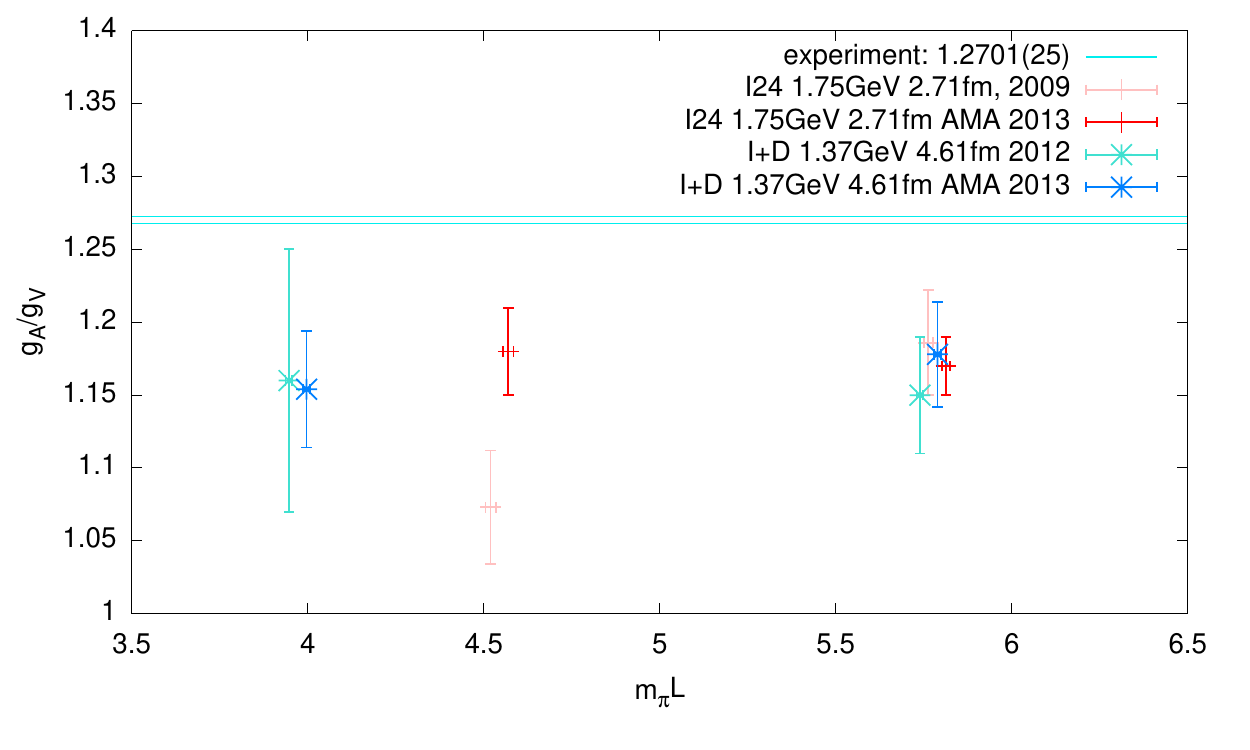}
\caption{\label{fig:gAgVmpi}
Dependence of the ratio, \(g_A/g_V\), of isovector axial charge, \(g_A\), and vector charge, \(g_V\), calculated with recent RBC+UKQCD 2+1-flavor dynamical DWF ensembles, on the pion mass squared, , \(m_\pi^2\) (left), and the finite-size scaling parameter, \(m_\pi L\) (right).
Solid symbols are the present results, while the faded ones are from our earlier publications.
While the precision agreement of the two ensembles at \(m_\pi L\) of about 5.8 has increased its significance (right),
the dependence on \(m_\pi^2\) does not show any sign of approaching the experiment as the pion mass squared decreases toward physical value (left).
}
\end{figure}

Let us first summarize the dependence on pion mass, \(m_\pi\).
In fig.\ \ref{fig:gAgVmpi} we plot our recent calculations of \(g_A/g_V\) against pion mass squared, \(m_\pi^2\), on the left, and against the finite-size scaling parameter, \(m_\pi L\), on the right.
The solid and faded red symbols denote results from the ensembles 1 and 2, and solid and faded blue from the ensembles 3 and 4.
The solid symbols are with the enhanced statistics this year, and faded ones are from our earlier calculations.

On the left we note the latest four calculations with improved statistics show no sign of approaching the experiment as the pion mass is set light, down to as light as about 171 MeV \cite{Arthur:2012opa}.
This behavior does not motivate us to fit them for any pion mass dependence.
About 10-\% deficit from the experiment is persistent down to this mass which is very close to physical.
We also note these results of ours are consistent with almost all the recent calculations of the quantity using various different actions, lattice spacings, and spatial volumes \cite{Lin:2012ev}.

On the right the agreement of the results from the ensembles 1 and 3, at \(m_\pi L \sim 5.8\), has increased in significance, from that of 1.19(4) and 1.15(5) a year ago to 1.17(2) and 1.18(4) now.
The solid points from the latest calculations also appear more smoothly monotonic in \(m_\pi L\).
This is not only because of the smaller statistical error from improved statistics, but also a change for ensemble 2, where the old result was calculated using the global currents but now we use the local currents:
The difference between the old and new here is consistent with the expected \(O(a^2)\) discretization effect.
All the other calculations are done with local currents only.

\section{No excited state seen}

An obvious question is why almost all the lattice calculations of \(g_A/g_V\) see the deficit of about 10 \%.
A focus of recent attention was whether the calculations suffer from excited-state contamination \cite{Capitani:2012gj,Lin:2012ev}.
The RBC collaboration was among the first to observe such contaminations in nucleon structure calculations, in some of the isovector structure function moments, namely the quark momentum and helicity fractions, \(\langle x \rangle_{u-d}\) and \(\langle x \rangle_{\Delta u-\Delta d}\) \cite{Lin:2008uz}.
However we have never detected such a contamination in our calculations of \(g_A/g_V\), probably because of our fairly well optimized nucleon source and sink smearing \cite{Lin:2012nv,Lin:2012jz}.
It is also likely the mixing matrix elements for the axial charge between the ground and excited states is smaller than for the structure function moments because the axial current is partially conserved: for conserved charge of the vector current the mixing matrix elements must vanish.
Nevertheless it was desired to more directly check this contamination.
With successful deployment of the AMA technique \cite{Blum:2012my,Blum:2012uh} we can now directly address this issue by comparing our calculations with different source-sink separations.
So far we have made this comparison with the ensemble 4 with the lightest of our pion mass of about 170 MeV, and compare the separations of 7 and 9 lattice units, or about 1.0 and 1.3 fm.
\begin{figure}
\includegraphics[width=0.48\textwidth]{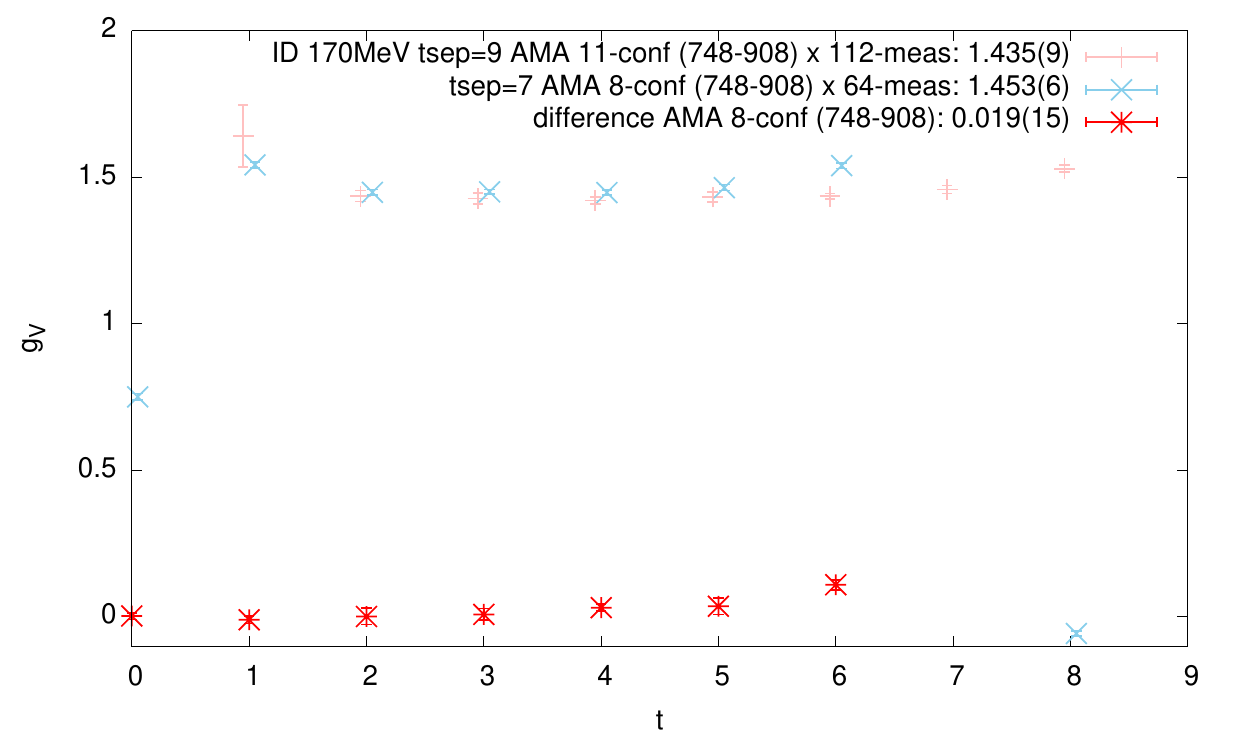}
\includegraphics[width=0.48\textwidth]{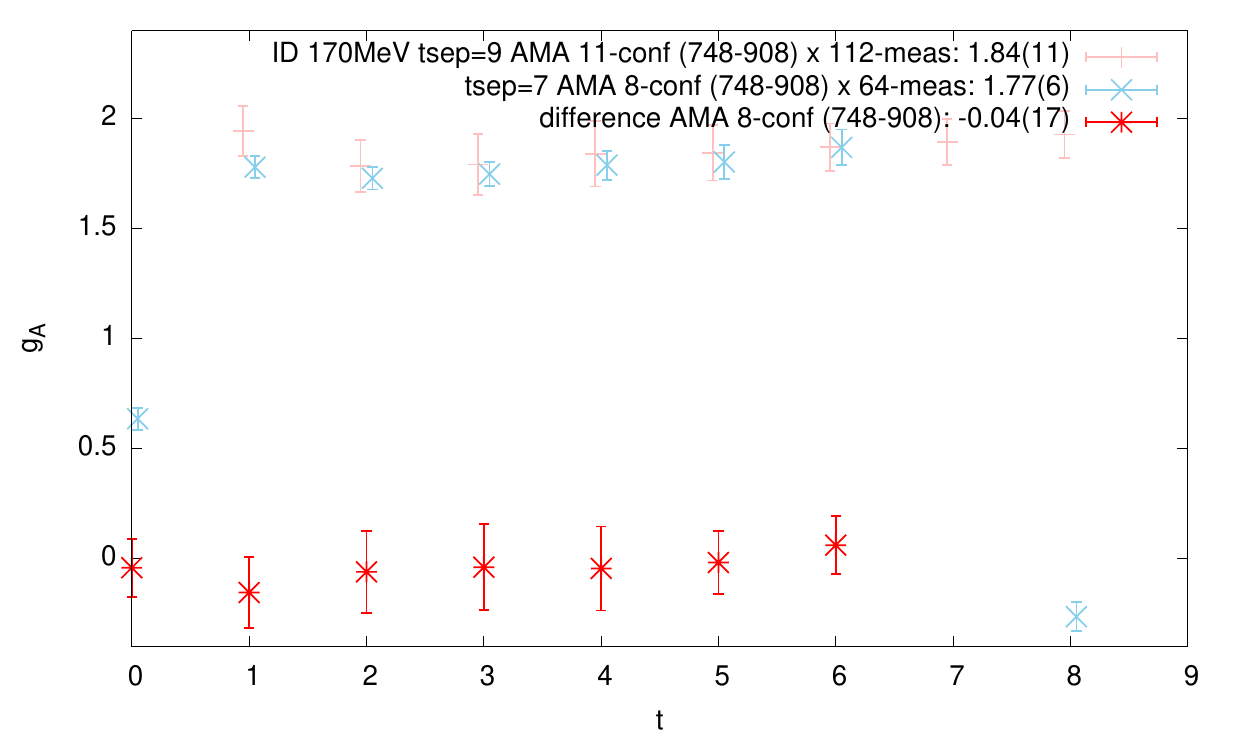}\\
\includegraphics[width=0.48\textwidth]{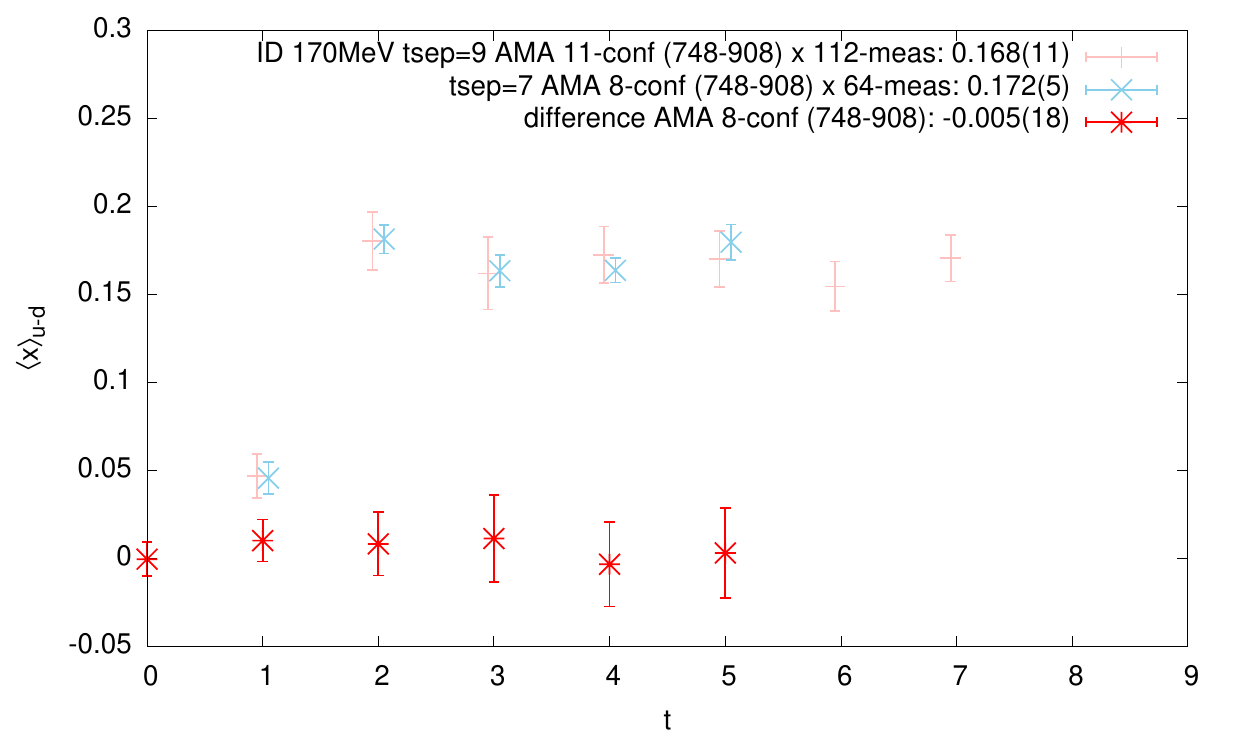}
\includegraphics[width=0.48\textwidth]{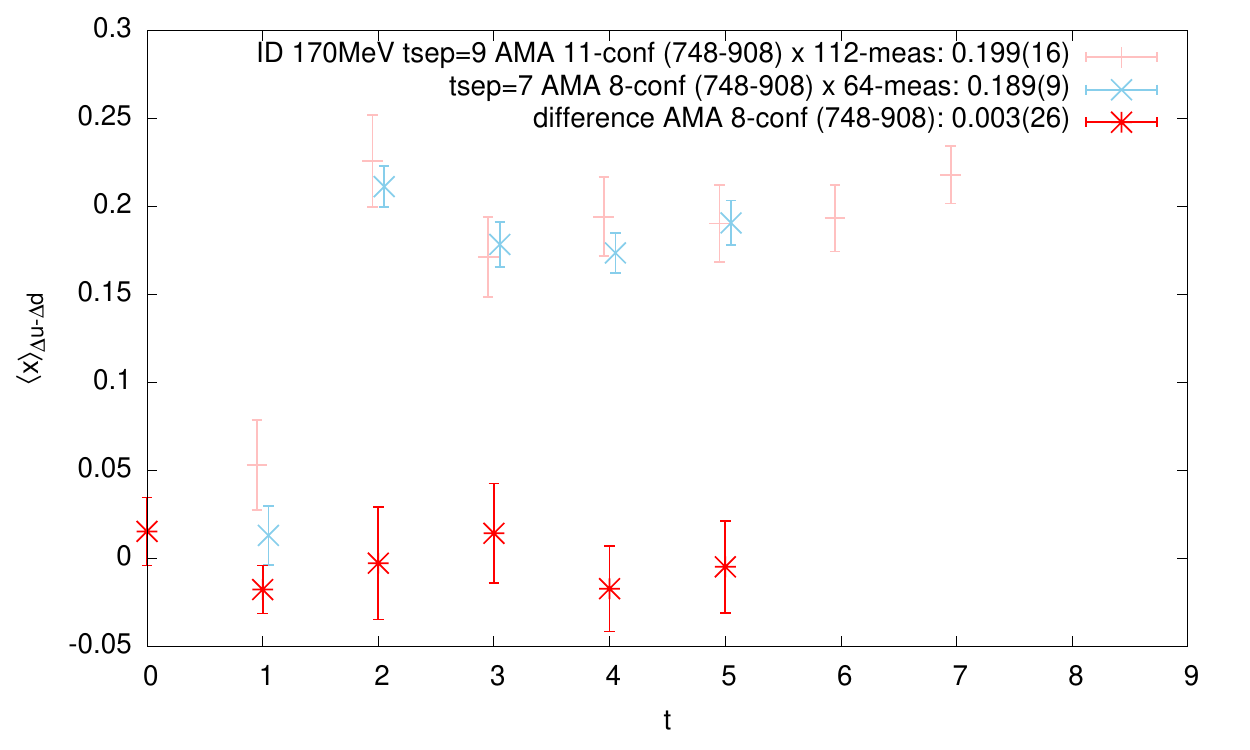}
\caption{\label{fig:NotExcited}
Comparison of two source-sink separations, 7 and 9 lattice units, or 1.0 and 1.3 fm:
Jack-knifed differences between the two separations (solid red symbols) fail to show any sign of excited-state contamination in any of the observables, the isovector vector charge, \(g_V\), axial charge, \(g_A\), quark momentum fraction, \(\langle x \rangle_{u-d}\), and helicity fraction \(\langle x \rangle_{\Delta u - \Delta d}\).
Faded symbols are the values of the observables themselves for the long (red) and short (blue) separations, respectively.
}
\end{figure}
In fig.\ \ref{fig:NotExcited} we present Jack-knife difference of the two source-sink separations for the four isovector quantities, the vector charge, \(g_V\), axial charge, \(g_A\), quark momentum fraction, \(\langle x \rangle_{u-d}\), and quark helicity fraction, \(\langle x \rangle_{\Delta u-\Delta d}\).
The solid red symbols are the differences for the eight configurations where both separation calculations exist.
The faded symbols are the values of the observables themselves in this range of trajectories, red with the longer and blue the shorter separations, respectively.

Of these the vector charge is diagonal among the mass eigen states and should be blind to the contamination.
It indeed is blind to a high accuracy even with relatively low statistics limited by the small number of samples, eight, currently available for the comparison.

For the other three observables there is no such protection so we should be able to see contamination of excited states if they are present.
As can be seen from fig.\ \ref{fig:NotExcited}, the differences are all consistent with zero:
We do not see such a contamination in our calculations.
This suggests our Gaussian smearing, with width of 6 lattice units, of the nucleon source and sink \cite{Lin:2012nv,Lin:2012jz} has effectively wiped out the excited states by the separation 7, or 1.0 fm, rather than the mixing matrix elements of the three observables conspire to vanish.

This observation of ours, that the excited-state contamination can be effectively controlled by smearing the source and sink appropriately, is consistent with many other groups who presented their nucleon structure calculations at Lattice 2013.

\section{Long-range autocorrelation in \(g_A\) but not in anything else}

As excited-state contamination is unlikely present in our calculations, we are yet to understand what other systematics can cause the observed deficit in calculated \(g_A/g_V\), of about 10 \%, compared with the experiment.
We have conjectured insufficient spatial volumes of the lattice may cause this \cite{Yamazaki:2008py}, based on scaling-like smooth and monotonic behavior of the quantity plotted against the finite-size scaling parameter, \(m_\pi L\).
Indeed the dependence on the lattice volume remains the least investigated systematics of these calculations, along with the lack of continuum limit.

\begin{figure}
\includegraphics[width=0.48\textwidth]{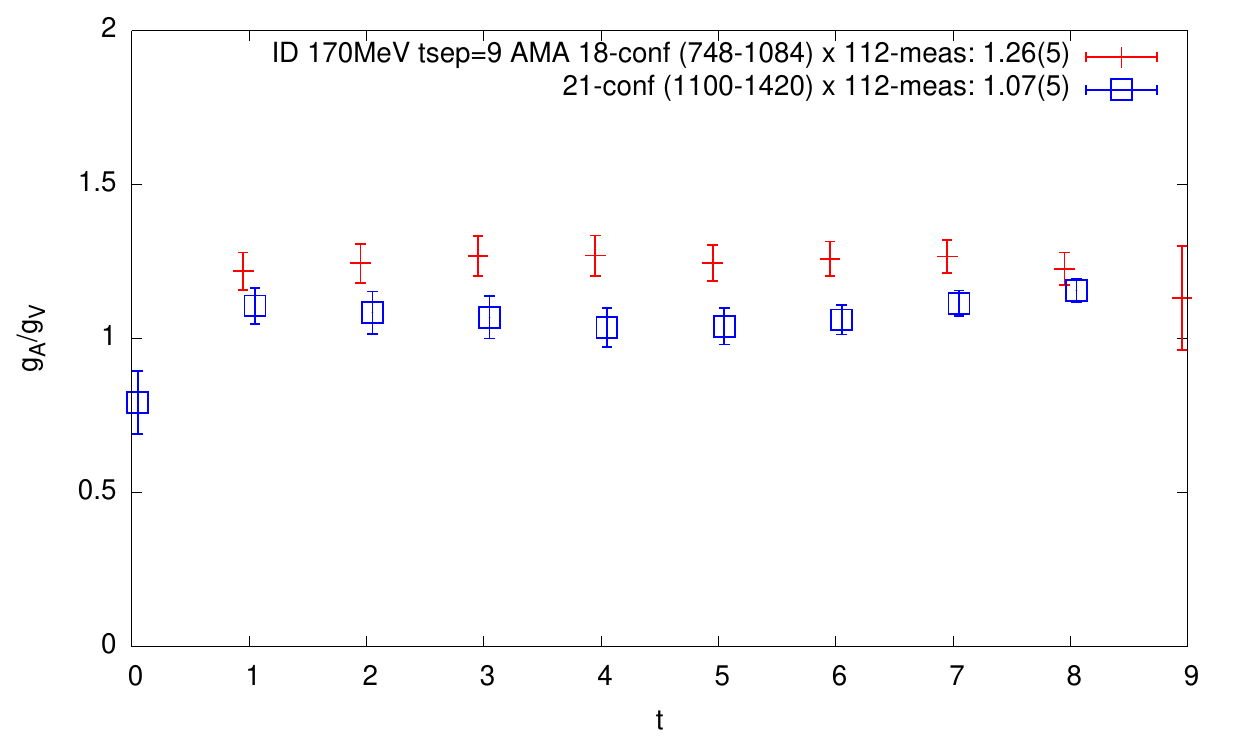}
\includegraphics[width=0.48\textwidth]{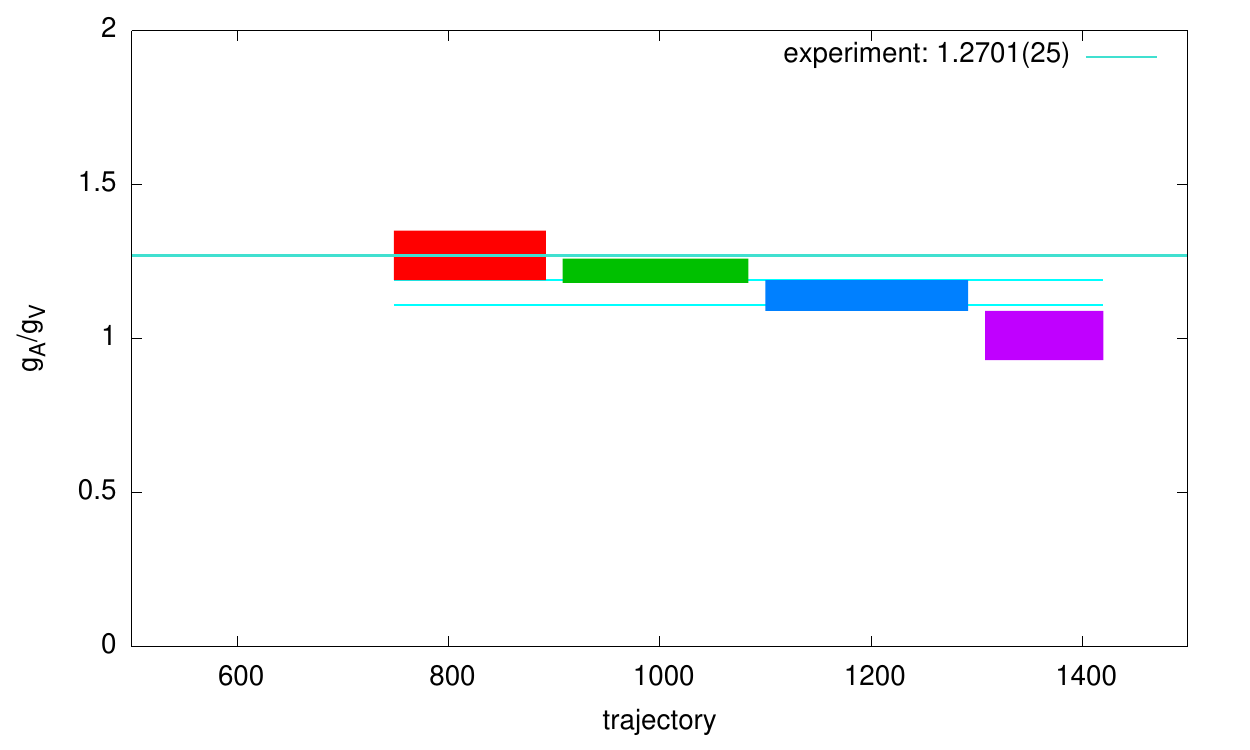}
\caption{\label{fig:gAgV2halves}
Very long-range autocorrelation is observed in both the isovector axial charge, \(g_A\), and its ratio, \(g_A/g_V\) to the vector charge in the lightest ensemble with \(m_\pi\) of 170 MeV.  On the left we present the plateaux of the ratio, \(g_A/g_V\), for the first (trajectory from 748 to 1084, red) and the second (1100 to 1420, blue) halves, respectively: fitted in the range from 2 to 7 lattice units, the values of1.26(5) for the first and 1.07(5) are almost four standard deviations away from each other.
On the right we present quarter-wise average along the hybrid MD time, from 748 to 892, 908 to 1084, 1100 to 1292, and 1308 to 1420: the values seem to drift monotonically from what is consistent with the experiment of 1.2701(25) in the first quarter to a value around 1.0 in the last quarter.
}
\end{figure}

Here we present a very long-range autocorrelation seen in the axial charge, \(g_A\), and its ratio to the vector charge, \(g_A/g_V\).
To see this we first divide the ensemble 4 into two halves, the first from the trajectory 748 to 1084 of the hybrid molecular dynamics evolution, and the second from 1100 to 1420.
With 16-trajectory sampling interval, there are 18 and 21 samples in these halves respectively, as some gaps exist in samplings due to computer malfunctions.
We had 112 sloppy AMA measurements for each of the samples.
In the left pane of fig.\ \ref{fig:gAgV2halves}
we compare the plateaux for \(g_A/g_V\) for these halves.
There is a big difference between the two, almost four standard deviations, with averages of 1.26(5) for the first and 1.07(5) for the second halves, respectively.
In the right pane we plot the averages calculated from four quarters, from trajectory 748 to 892, 908 to 1084, 1100 to 1292, and 1308 to 1420:
The quarter-wise average starts with a value consistent with experiment within a standard deviation in the first quarter, then drifts monotonically down to a value close to 1.0 in the last quarter.
There is a very long-range autocorrelation that extend our entire sample from trajectory 748 to 1420.

\begin{figure}
\includegraphics[width=0.48\textwidth]{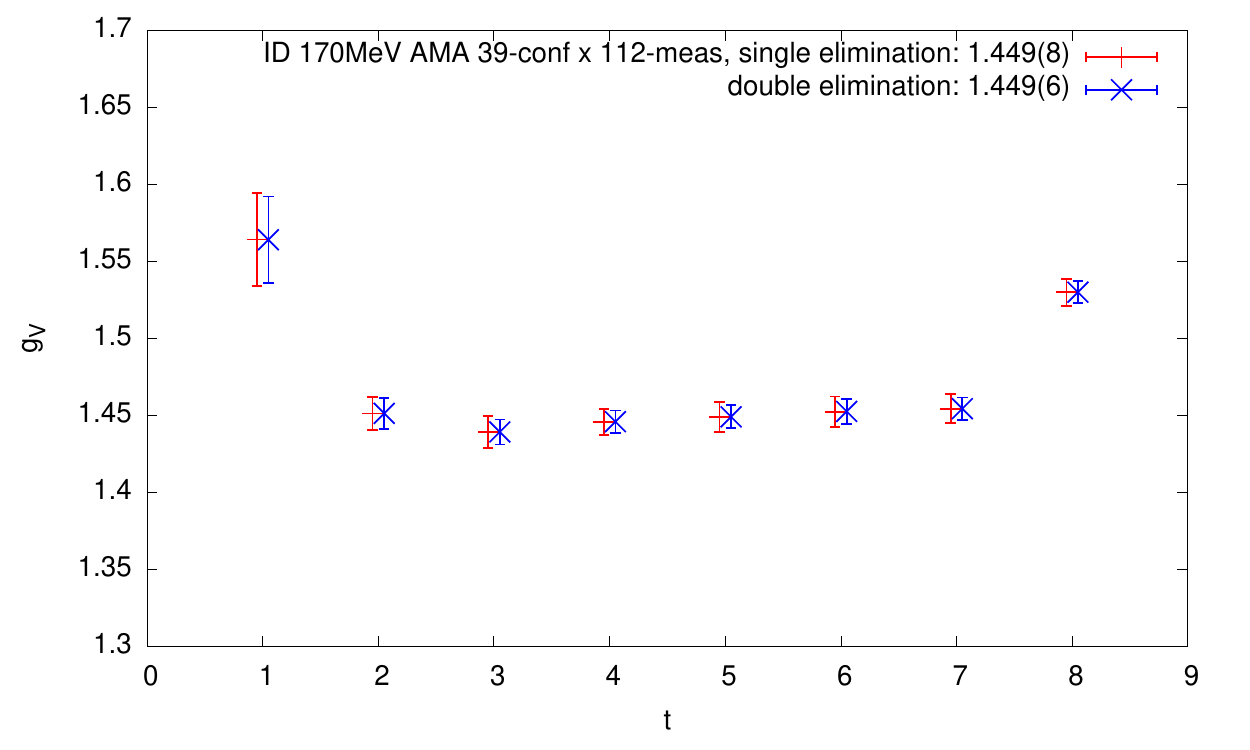}
\includegraphics[width=0.48\textwidth]{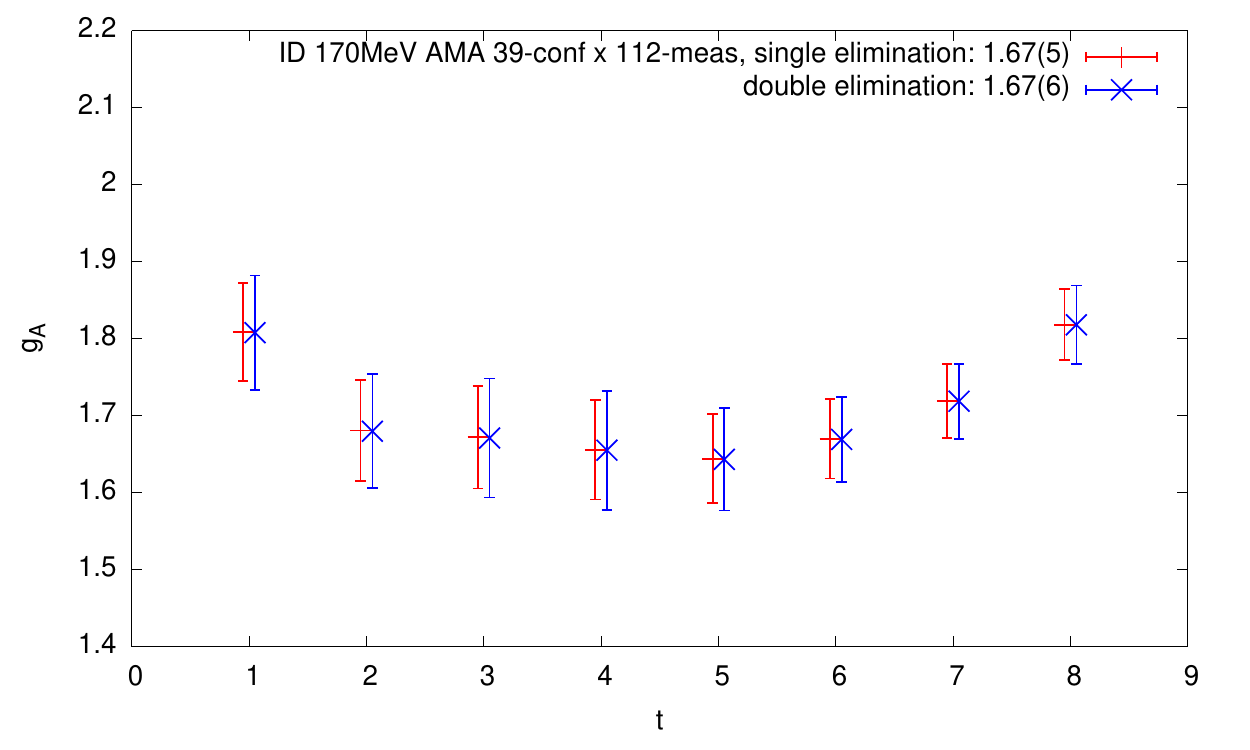}\\
\includegraphics[width=0.48\textwidth]{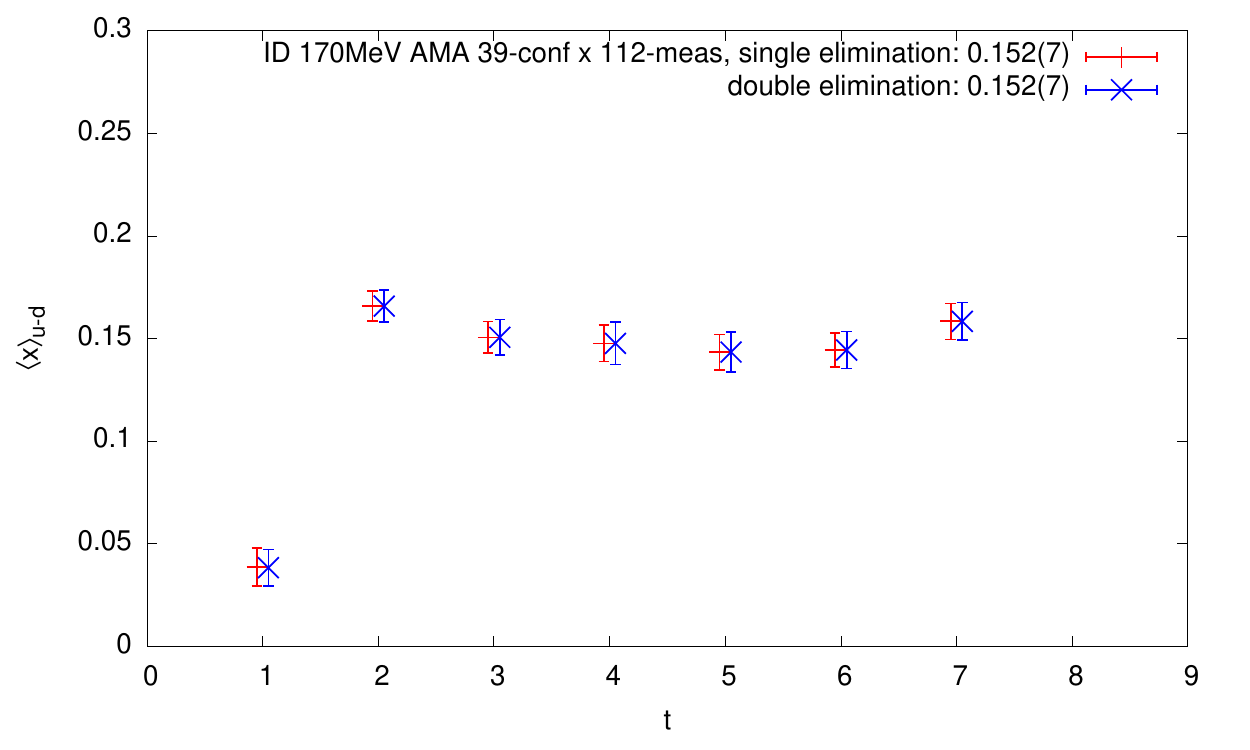}
\includegraphics[width=0.48\textwidth]{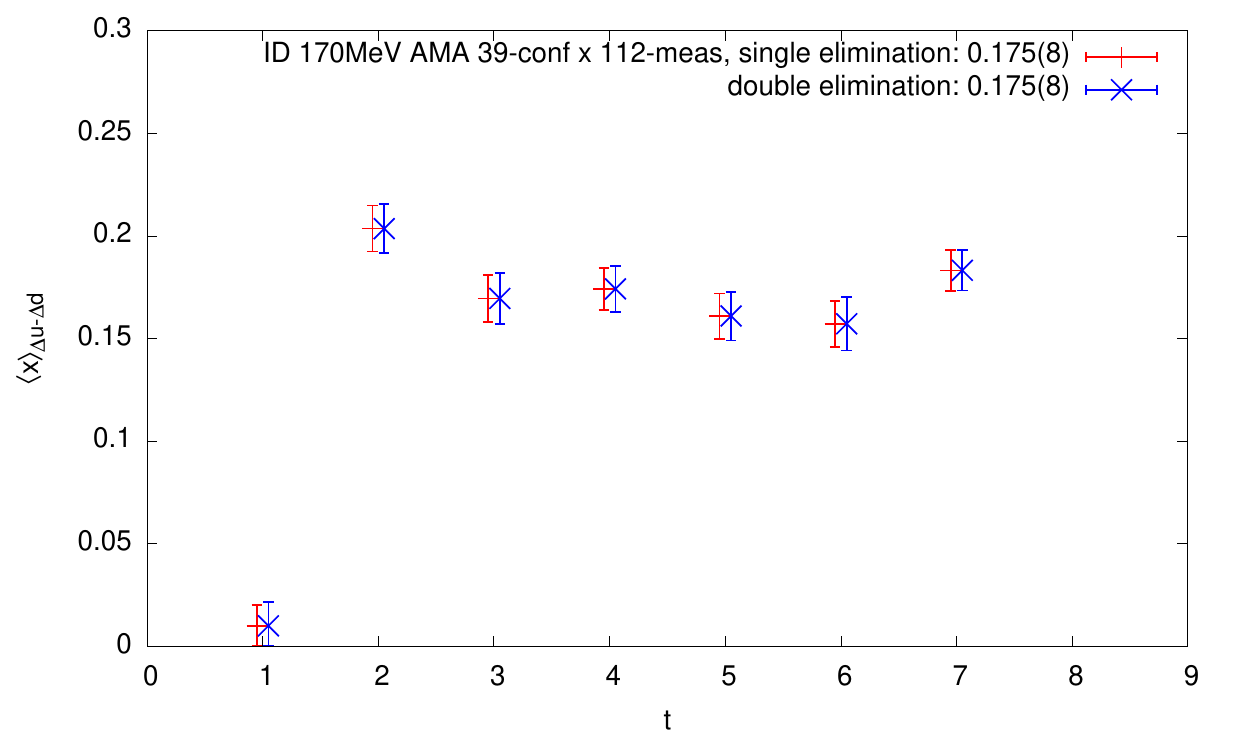}
\caption{\label{fig:blocking}
Blocked Jack-knife analyses for the four observables, \(g_V\) at upper left, \(g_A\) upper right, \(\langle x \rangle_{u-d}\) lower left, and \(\langle x \rangle_{\Delta u - \Delta d}\) lower right.
When two successive configurations are blocked together in the lightest ensemble with \(m_\pi\) of 170 MeV,
the statistical error grows significantly for the isovector axial charge, \(g_A\), and its ratio, \(g_A/g_V\) to the vector charge (not shown), in consistency with the long-range autocorrelation.
In contrast, no other observable shows such growth.
}
\end{figure}

In contrast, no such autocorrelation is observed in the other three observables, the isovector vector charge, \(g_V\),  quark momentum fraction, \(\langle x \rangle_{u-d}\), and quark helicity fraction, \(\langle x \rangle_{\Delta u-\Delta d}\).
In fig. \ref{fig:blocking}
we present the results of a blocking analysis where two consecutive samples are blocked.
The statistical errors for the axial charge, \(g_A\), grows almost by a factor of \(\sqrt{2}\), indicating very strong correlation for two successive configurations, as is expected from the slow decorrelation seen in fig. \ref{fig:gAgV2halves}.
But no such growth of statistical error is seen in any of the other three observables, indicating the two successive samples separated by 16 hybrid MD time are independent of each other.

A way to interpret this very long-range autocorrelation seen in the axial charge but not in other observables is that the lattice volume is insufficient for the former which couples strongly with the pion and would be dominated by the pion pole as pion mass is set light \cite{Adkins:1983ya}.
Nucleon is hardly point-like in terms of its coupling with pion:
How does this reconcile with the conventional nuclear models?

\section{Conclusions}

RBC and UKQCD collaborations continue to work on nucleon structure using their 2+1-flavor dynamical DWF ensembles, in particular
\begin{enumerate}
\item Iwasaki gauge action, inverse lattice spacing, \(a^{-1}\), of about 1.7 GeV, lattice linear extent, \(L\), of about 2.7 fm, pion mass, \(m_\pi\) of about 420 MeV,
\item same as the above except \(m_\pi\) of about 330 MeV,
\item Iwasaki \(\times\) dislocation suppressing determinant ratio (DSDR) gauge action, \(a^{-1}\) of about 1.4 GeV, \(L\) of about 4.6 fm, \(m_\pi\) of about 250 MeV,
\item same as the above except \(m_\pi\) of about 170 MeV.
\end{enumerate}
We enhanced statistics for all of these, by adding more source-sink pairs for 1 and 3, and by employing the AMA technique for 2 and 4, resulting in very solid signals for isovector observables such as vector charge,  \(g_V\), axial charge, \(g_A\), quark momentum fraction, \(\langle x \rangle_{u-d}\), and quark helicity fraction, \(\langle x \rangle_{\Delta u - \Delta d}\).
We found:
\begin{itemize}
\item By comparing two source-sink separations of about 1.0 and 1.3 fm for the ensemble 4, we established absence of excited-state contamination there.
\item We see persistent deficit in the ratio, \(g_A/g_V\), of the isovector axial and vector charges compared with the experiment in all the four ensembles, with increased statistical significance for all of them.
\item A very long-range autocorrelation is seen in the axial charge, \(g_A\), and its ratio to the vector charge, \(g_A/g_V\), in the ensemble 4, and possibly in 2 as well, suggesting insufficient volume.
\item No such long-range autocorrelation is seen for the other three observables, the isovector vector charge, \(g_V\), quark momentum fraction, \(\langle x \rangle_{u-d}\), and quark helicity fraction, \(\langle x \rangle_{\Delta u - \Delta d}\).
\end{itemize}
We plan to enhance the statistics further at least for the ensemble 4, by extending the AMA sampling beyond the trajectory 1420 through 1920, so we better understand the long-range autocorrelation in the axial charge.
We are also starting a collaboration with LHP to investigate nucleon structure at physical mass using the RBC+UKQCD 2+1-flavor dynamical DWF ensemble with Iwasaki gauge at inverse lattice spacing of about 1.75 GeV and spatial extent of about 5.4 fm.

I thank Tom Blum, Taku Izubuchi, Chulwoo Jung, Meifeng Lin and Eigo Shintani for close collaboration on the work presented here, and past and present members of the RBC and UKQCD collaborations who contributed to the four 2+1-flavor dynamical DWF ensembles.
The ensembles were generated using four QCDOC computers of Columbia University, Ediburgh University, RIKEN-BNL Research Center (RBRC) and  USQCD collaboration at Brookhaven National Laboratory, and a Bluegene/P computer of Argonne Leadership Class Facility (ALCF) of Argonne National Laboratory provided under the INCITE Program of US DOE.
Calculations of nucleon observables were done using RIKEN Integrated Cluster of Clusters (RICC) at RIKEN, Wako, and various Teragrid and XSEDE clusters of US NSF.


\begin{thebibliography}{99}

\bibitem{Sasaki:2003jh} 
  S.~Sasaki {\it et al.},  
  Phys.\ Rev.\ D {\bf 68}, 054509 (2003).

\bibitem{Orginos:2005uy} 
  K.~Orginos, T.~Blum and S.~Ohta,
  Phys.\ Rev.\ D {\bf 73}, 094503 (2006).

\bibitem{Lin:2008uz} 
  H.~-W.~Lin, T.~Blum, S.~Ohta, S.~Sasaki and T.~Yamazaki,
  Phys.\ Rev.\ D {\bf 78}, 014505 (2008).
    
\bibitem{Yamazaki:2008py} 
  T.~Yamazaki {\it et al.},  
  Phys.\ Rev.\ Lett.\  {\bf 100}, 171602 (2008).
  
\bibitem{Yamazaki:2009zq} 
  T.~Yamazaki, Y.~Aoki, T.~Blum, H.~-W.~Lin, S.~Ohta, S.~Sasaki, R.~Tweedie and J.~Zanotti,
  Phys.\ Rev.\ D {\bf 79}, 114505 (2009).
    
\bibitem{Aoki:2010xg} 
  Y.~Aoki, T.~Blum, H.~-W.~Lin, S.~Ohta, S.~Sasaki, R.~Tweedie, J.~Zanotti and T.~Yamazaki,
  Phys.\ Rev.\ D {\bf 82}, 014501 (2010).
      
\bibitem{Lin:2012nv} 
  M.~Lin and S.~Ohta,  
  PoS LATTICE {\bf 2012}, 171 (2012).

\bibitem{Lin:2012jz} 
  M.~Lin, 
  PoS LATTICE {\bf 2012}, 172 (2012).

\bibitem{Aoki:2010dy} 
  Y.~Aoki {\it et al.}, 
  Phys.\ Rev.\ D {\bf 83}, 074508 (2011)
  
\bibitem{Arthur:2012opa} 
  R.~Arthur {\it et al.}, 
  Phys.\ Rev.\ D {\bf 87}, 094514 (2013).

\bibitem{Blum:2012my} 
  T.~Blum, T.~Izubuchi and E.~Shintani,
  PoS LATTICE {\bf 2012}, 262 (2012).
  
\bibitem{Blum:2012uh} 
  T.~Blum, T.~Izubuchi and E.~Shintani,
  [arXiv:1208.4349 [hep-lat]].

\bibitem{Lin:2012ev} 
  H.~-W.~Lin,
  PoS LATTICE {\bf 2012}, 013 (2012);
  See also a review by S.~Syritsyn, in these proceedings.
  
\bibitem{Capitani:2012gj} 
  S.~Capitani, M.~Della Morte, G.~von Hippel, B.~Jager, A.~Juttner, B.~Knippschild, H.~B.~Meyer and H.~Wittig,
  Phys.\ Rev.\ D {\bf 86}, 074502 (2012).

\bibitem{Adkins:1983ya} 
  G.~S.~Adkins, C.~R.~Nappi and E.~Witten,
  Nucl.\ Phys.\ B {\bf 228}, 552 (1983).

\end{thebibliography}
\end{document}